\documentclass[twocolumn,showpacs,preprintnumbers,amsmath,amssymb]{revtex4}
\usepackage{graphicx}
\usepackage{dcolumn}
\usepackage{bm}

\def\mop#1{\mathop{\rm #1}\nolimits}

\def\Im{\mop{Im}}
\def\tr{\mop{tr}}

\begin{document}

\preprint{PUPT-2266}

\title{Thermodynamics and bulk viscosity of approximate black hole duals to finite temperature quantum chromodynamics}

\author{Steven S. Gubser}\email{ssgubser@Princeton.EDU}
\author{Abhinav Nellore}\email{anellore@Princeton.EDU}
\author{Silviu S. Pufu}\email{spufu@Princeton.EDU}
\author{F\'abio D. Rocha}\email{frocha@Princeton.EDU}
\affiliation{Joseph Henry Laboratories, Princeton University, Princeton, NJ 08544}

\date{April 2008}

\begin{abstract}
We consider classes of translationally invariant black hole solutions whose equations of state closely resemble that of QCD at zero chemical potential.  We use these backgrounds to compute the ratio $\zeta/s$ of bulk viscosity to entropy density.  For a class of black holes that exhibits a first order transition, we observe a sharp rise in $\zeta/s$ near $T_c$.  For constructions that exhibit a smooth cross-over, like QCD does, the rise in $\zeta/s$ is more modest.  We conjecture that divergences in $\zeta/s$ for black hole horizons are related to extrema of the entropy density as a function of temperature.
\end{abstract}


\maketitle

The anti-de Sitter / conformal field theory (AdS/CFT) correspondence \cite{Maldacena:1997re, Gubser:1998bc, Witten:1998qj} has generated interest in using thermal $\mathcal{N}=4$ super-Yang-Mills theory (SYM) to understand quantum chromodynamics (QCD) at finite temperature.  A conspicuous shortcoming of this approach is precisely the con\-formal invariance of SYM\@.  This implies, for instance, that in the SYM plasma, the speed of sound $c_s$ equals $1/\sqrt{3}$ and that the bulk viscosity $\zeta$ vanishes at all temperatures.  QCD only exhibits conformal behavior in the high-temperature regime.  In order to roughly capture the behavior of QCD across a larger range of temperatures, we are led to consider gravity duals of gauge theories that break conformal invariance \footnote{There is a large literature on ``AdS/QCD,'' in which models of QCD are based on holographic duals of theories with broken conformal invariance, similar to our constructions.  The AdS/QCD literature includes relatively little investigation of thermodynamic properties, focusing instead on the vacuum and hadron spectroscopy; see however \cite{Benincasa:2006ei,Kajantie:2006hv,Andreev:2007zv}.}.  The minimal action on the gravity side that can describe Lorentz-invariant, non-conformal theories is
 \begin{equation}
    S={1 \over 2 \kappa_5^2} \int d^5 x\sqrt{-g}
      \left(R-{1 \over 2}(\partial\phi)^2 - V(\phi)\right) \,,
        \label{Action}
 \end{equation}
where $V(\phi)$ is the potential for the bulk scalar field $\phi$, and $\kappa_5 = \sqrt{8 \pi G_5}$ is the five-dimensional gravitational constant.  We restrict our attention to backgrounds of the type
 \begin{align}
   ds^2 &= e^{2 A(r)} \left(-h(r) dt^2 + d\vec{x}^2 \right) + e^{2 B(r)} {dr^2\over h(r)} \label{MetricAnsatz} \\
   \phi &= \phi(r) \qquad d\vec{x}^2 \equiv (dx^1)^2 + (dx^2)^2 + (dx^3)^2 \,. \label{PhiAnsatz}
 \end{align}
This form is the most general ansatz with translational symmetry in the $(t, \vec{x})$ coordinates and $SO(3)$ symmetry in the $\vec{x}$ directions, as is appropriate to describe an infinite static thermal plasma.  The equations of motion for the functions $A$, $B$, and $h$ come from plugging the ansatz \eqref{MetricAnsatz}--\eqref{PhiAnsatz} into the equations of motion following from the action \eqref{Action}.  See \cite{Gubser:2008ny}, in which techniques for finding solutions of the form \eqref{MetricAnsatz}--\eqref{PhiAnsatz} and the corresponding equation of state are more fully explained.

The $AdS_5$-Schwarzschild solution can be recovered as the limit of the above construction where $\phi$ vanishes and $V(\phi)$ is a negative cosmological constant term.  More generally, if at small $\phi$ one has
 \begin{equation}
   V(\phi) = -{12\over L^2} + {1\over 2} m^2 \phi^2 + {\cal O}(\phi^4)\,, \label{SmallPhi}
 \end{equation}
then the gravity solution \eqref{MetricAnsatz}--\eqref{PhiAnsatz} will be asymptotic to anti-de Sitter space with radius $L$.  An asymptotically AdS spacetime on the gravity side is equivalent to conformal invariance of the field theory in the UV\@.  Gravity backgrounds constructed from potentials which satisfy \eqref{SmallPhi} are dual to relevant deformations of a conformal field theory:
 \begin{equation}
   {\cal L} = {\cal L}_{\rm CFT} + \Lambda_\phi^{4-\Delta} {\cal O}_{\phi} \,, \label{FTLagrangian}
 \end{equation}
where $\Lambda_\phi$ is the energy scale of the deformation and $\Delta$ is the dimension of the operator ${\cal O}_\phi$ dual to $\phi$.  According to the AdS/CFT dictionary, $\Delta$ can be identified with the larger root of
 \begin{equation}
   \Delta(\Delta - 4) = m^2 L^2 \,. \label{DeltaEQ}
 \end{equation}
We will only be interested in the case $2 < \Delta < 4$, which corresponds to relevant deformations that obey the Breitenlohner-Freedman (BF) bound \cite{Breitenlohner:1982bm, Breitenlohner:1982jf, Mezincescu:1984ev}.

A background of the form \eqref{MetricAnsatz}--\eqref{PhiAnsatz} has an event horizon if $h$ has a zero.  Let $r_H$ be the value of $r$ closest to the conformal boundary where $h$ vanishes.  Thermodynamic quantities such as entropy density $s$ and temperature $T$ are parameterized by $r_H$:
 \begin{equation}
   s = {2\pi \over \kappa_5^2} e^{3A(r_H)} \qquad
   T = {e^{A(r_H) - B(r_H)} |h'(r_H)| \over 4\pi} \label{EntropyAndTemperature} \,.
 \end{equation}
The speed of sound $c_s$ can be computed from
 \begin{equation}
   c_s^2 = {d \log T \over d \log s} \,. \label{csDef}
 \end{equation}
We exclude from consideration nonzero chemical potential for baryon number.  To include this, we would have to add a gauge field to the action \eqref{Action} and consider charged black holes.

If $V = V_0 e^{\gamma\phi}$ with $V_0 < 0$, then the equations of motion following from \eqref{Action} can be solved analytically \cite{Chamblin:1999ya}, and the speed of sound is constant: $c_s^2 = {1 \over 3} - {\gamma^2 \over 2}$.  But these black holes aren't asymptotically anti-de Sitter because $V(\phi)$ has no maximum.  If instead $V(\phi)$ interpolates smoothly between \eqref{SmallPhi} for small $\phi$ and $V_0 e^{\gamma\phi}$ for large $\phi$, then the black hole solutions have a temperature-dependent speed of sound, $c_s(T)$.  In \cite{Gubser:2008ny}, the mapping between $V(\phi)$ and $c_s(T)$ is explored in some detail.  Within certain limits, given $c_s(T)$, one can find a $V(\phi)$ to reproduce it using black holes.

It probably isn't possible to obtain an arbitrary $V(\phi)$ from string theory.  However, it is typical in gauged supergravity to find potentials with local extrema and exponential increase or decrease as canonically normalized scalars become large.  In any case, it is our goal to design a potential $V(\phi)$ to reproduce the equation of state of QCD\@.

\begin{figure*}
  \centering{\includegraphics[width=3.3in]{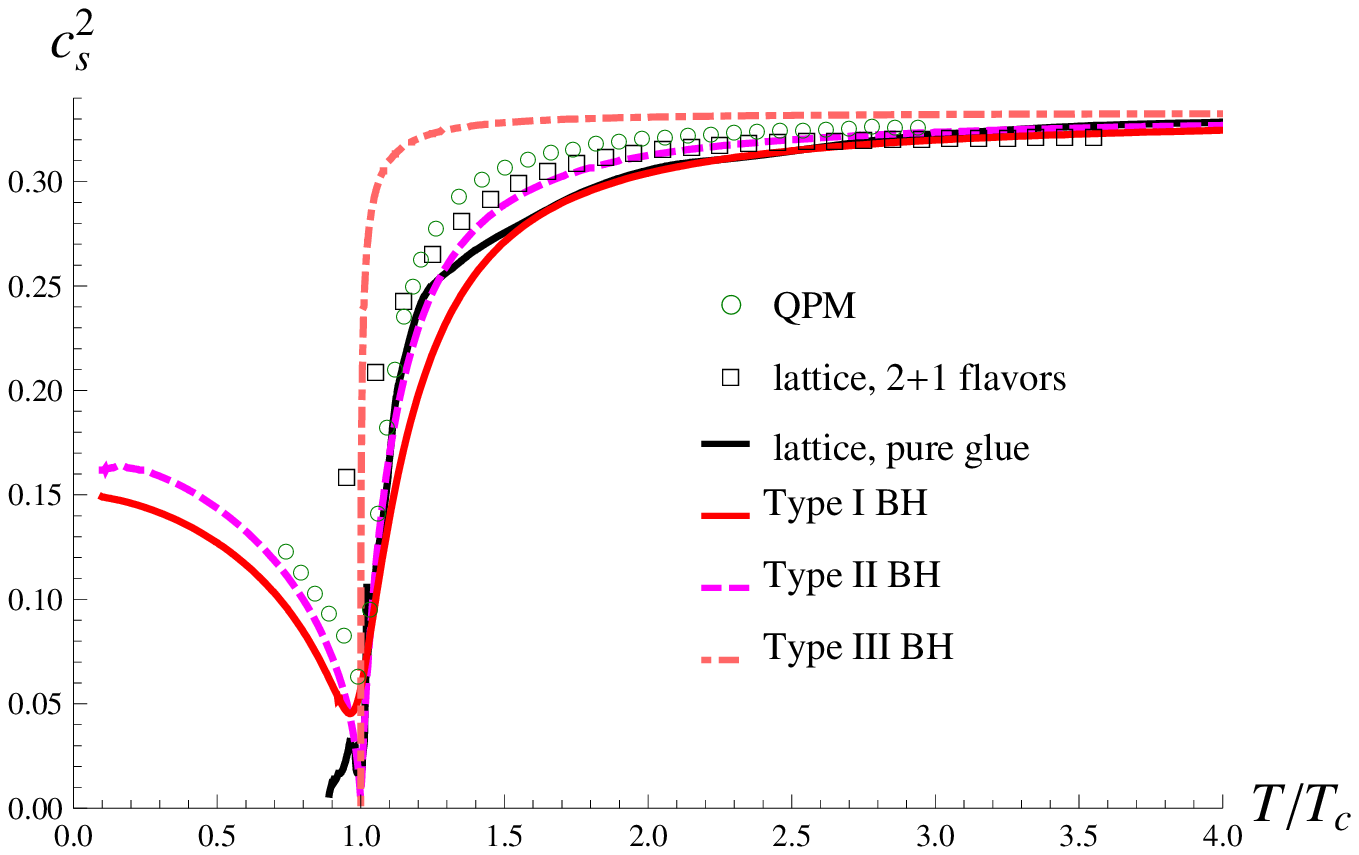}\hfill\includegraphics[width=3.3in]{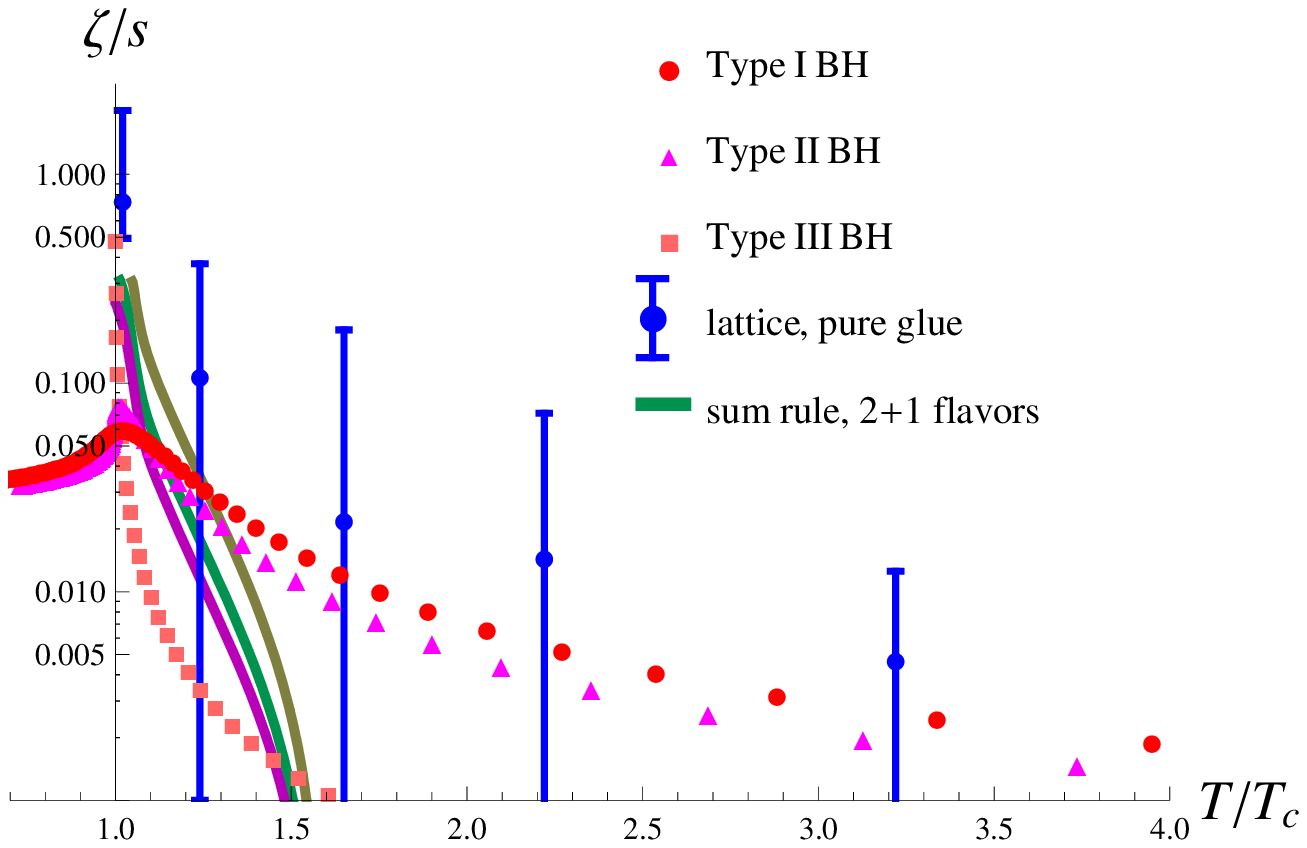}}
  \caption{LEFT: A comparison of different $c_s^2(T)$ curves.  The solid red curve corresponds to the potential \eqref{FirstPotential} (Type I black holes).  The dashed magenta curve corresponds to the potential \eqref{TunedV} (Type II black holes).  The dot-dashed orange curve corresponds to the potential in \eqref{ConfinementPotential} (Type III black holes) with $a= 1$ and $b$ adjusted so that the dimension of the operator dual to $\phi$ is $\Delta \approx 3.93$.  We also show lattice results for pure glue (solid black curve) and $2+1$-flavor QCD (open black squares), as well as $c_s^2(T)$ for a $2+1$-flavor quasiparticle model (QPM)\@. The pure glue curve is based on \cite{Boyd:1996bx} and private communications from F.~Karsch.  The $2+1$-flavor lattice QCD points are based on \cite{Cheng:2007jq}; for these points, we take $T_c=187 \, {\rm MeV}$, as estimated from the halfway point of the initial rise in an $(\epsilon-3p)/T^4$ curve from \cite{Cheng:2007jq}.  This differs from the value $T_c = 196(3) \, {\rm MeV}$ computed by a different method in \cite{Cheng:2007jq}.  The QPM points are based on \cite{Bluhm:2007nu}.\\
    RIGHT: A comparison of $\zeta/s$ results for the black holes described above (see legend) and the results of \cite{Meyer:2007dy,Kharzeev:2007wb}.  Lattice results for pure glue from \cite{Meyer:2007dy} are shown in blue.  The solid curves correspond to the sum rule result for QCD with $2+1$ flavors from \cite{Kharzeev:2007wb}, using three representative values of the frequency parameter $\omega_0$.}
  \label{EOSandZetaComparison}
\end{figure*}

It is perhaps surprising that the simple potential
 \begin{equation}
   V(\phi) = {-12 \cosh{\gamma \phi} + b \phi^2 \over L^2} \label{FirstPotential}
 \end{equation}
with $\gamma \approx 0.606$ and $b \approx 2.057$ approximately reproduces the squared speed of sound versus temperature as derived from lattice data on $2+1$-flavor QCD: see figure~\ref{EOSandZetaComparison}.  Because the equation of state exhibits a cross-over rather than a sharp phase transition, we have to use some prescription to determine $T_c$ in order to plot $c_s^2$ versus $T/T_c$.  We define $T_c$ as the inflection point of $s/T^3$ as a function of $T$.

The quoted value for $\gamma$ corresponds to setting $c_s^2 = 0.15$ in the extreme IR.  Although hadron resonance gas models give values of $c_s^2$ ranging as high as $0.2$ (see \cite{Bluhm:2007nu} and references therein), the value $c_s^2=0.15$ is in a phenomenologically interesting range.  The equation of state following from \eqref{FirstPotential} is fairly close to the quasiparticle model of \cite{Bluhm:2007nu}, based on a chiral extrapolation of lattice data.  The black hole model is complementary to a quasiparticle description in that it should work well precisely when no weakly coupled quasiparticle description is available, reminding us of the correspondence principle of \cite{Horowitz:1996nw}.  Thus, the picture we advocate in using the potential \eqref{FirstPotential} is that the approximate validity of a black hole description of QCD is not lost suddenly during the smooth cross-over, but instead gradually, so that the black hole continues to give an approximate guide to the dynamics at least down to $T_c$, and perhaps even somewhat below it.  This is a departure from a more traditional picture, inspired in part by large $N$ counting, where there is a sharp transition (usually first order) between a black hole description of a deconfined phase and a horizon-free description of the confined phase: see for example \cite{Witten:1998zw}.

The quoted value for $b$ corresponds to setting $\Delta \approx 3.93$, which is the dimension of $\tr F_{\mu\nu}^2$ in QCD computed at three loops and energy scale $Q = 3\,{\rm GeV}$.  Here, $F_{\mu\nu}$ is the rescaled field strength that appears in the QCD lagrangian
 \begin{equation}
   {\cal L}_{\rm QCD} = -{1\over 8 \pi \alpha_0} \tr F_{\mu\nu}^2 + \textrm{fermionic terms}\,,  \label{QCDLagrangian}
 \end{equation}
where $\alpha_0 = g_0^2/4 \pi$ is the bare strong coupling constant.  To compute the dimension of $\tr F_{\mu\nu}^2$, one starts by noticing that the trace of the QCD stress-energy tensor
 \begin{equation}
   T_\mu^{\phantom{\mu}\mu} = {\beta(\alpha) \over 8 \pi \alpha^2} \tr F_{\mu\nu}^2 + \textrm{fermionic terms} \label{GotTQCD}
 \end{equation}
is RG-invariant, so it should scale classically.  In \eqref{GotTQCD}, $\alpha$ is the renormalized coupling at scale $Q$, and $\beta(\alpha)$ is the QCD beta-function
 \begin{equation}
   \beta(\alpha) = Q {d \alpha \over d Q} \,.  \label{BetaDef}
 \end{equation}
For any operator ${\cal O}$, $d{\cal O}/d\log Q = - {\cal O} \Delta$, where $\Delta$ is the sum of the classical and anomalous dimensions of ${\cal O}$.  Thus, differentiating \eqref{GotTQCD} with respect to $\log Q$, one obtains \footnote{This argument was suggested to one of us by A.~Polyakov.}
 \begin{equation}
   \Delta = 4 + \beta'(\alpha) - {2 \beta(\alpha) \over \alpha} \,. \label{GotDelta}
 \end{equation}
Reference \cite{Yao:2006px} contains the exact expressions for $\beta(\alpha)$ and $\alpha(Q)$ \footnote{There is a factor of $2$ difference between our definition of $\beta(\alpha)$ and the quantity that appears in the middle equality in (9.4$a$) of \cite{Yao:2006px}.}.  At $Q = 3\,{\rm GeV}$, we get $\alpha \approx 0.253$ and $\Delta \approx 3.93$ at three loops.

In summary, the UV matching to QCD doesn't attempt to capture asymptotic freedom, which probably requires going beyond the supergravity approximation; instead we match onto QCD at a finite scale, above which asymptotic freedom is replaced by conformal invariance.

Black holes constructed with the potential \eqref{FirstPotential} may slightly underestimate the rapidity of the cross-over of QCD\@.  We therefore consider an alternative potential,
 \begin{equation}\label{TunedV}
  V(\phi) = {-12 \cosh{\gamma \phi} + b_2 \phi^2 + b_4 \phi^4 + b_6 \phi^6 \over L^2}\,,
 \end{equation}
with $\gamma \approx 0.606$, $b_2 \approx 1.975$, $b_4 \approx -0.030$, and $b_6 \approx -0.0004$, where the specific values are chosen so as to sharpen the cross-over almost to a second order phase transition.  One can see from figure~\ref{EOSandZetaComparison} that $c_s^2$ for the choice \eqref{TunedV} is a close match to pure glue data of \cite{Boyd:1996bx} for $T>T_c$, so its behavior close to $T_c$ is probably sharper than QCD's.  But, by design, it still has a behavior reminiscent of hadron gas phenomenology for $T<T_c$.

The shear viscosity of all black hole solutions we construct satisfies $\eta/s = 1/4\pi$ \cite{Buchel:2003tz} because we exclude higher derivative terms from the action.  This low value of $\eta$ reminds us that the regime of validity of a black hole description cannot extend too far above $T_c$ or too far below it.  The bulk viscosity can also be studied (see for example \cite{Parnachev:2005hh,Benincasa:2005iv,Benincasa:2006ei,Buchel:2007mf}), and it is particularly interesting to inquire how it behaves near $T_c$.  There is a proposal \cite{Karsch:2007jc} that QCD exhibits a sharp rise in $\zeta/s$ close to the deconfinement transition, signaling ``soft statistical hadronization'' of the QGP\@.  See also the earlier works \cite{Kharzeev:2007wb, Meyer:2007dy}, which deal with pure glue.

Bulk viscosity can be computed from the Kubo formula
 \begin{align}
   \zeta &= {1\over 9} \lim_{\omega \to 0} {1\over \omega} \Im G_R(\omega) \\
   G_R(\omega) &\equiv \int d^3x\, dt\, e^{i \omega t} \theta(t) \langle [T_{ii}(t, \vec{x}), T_{kk}(0, 0)] \rangle\,, \label{Kubo}
 \end{align}
where $G_R(\omega)$ is the retarded two-point function of the spatial trace $T_{ii}$ of the stress-energy tensor.  Two-point functions of the stress-energy tensor can be computed within AdS/CFT by examining the metric perturbations $\delta g_{ij}$ around the background \eqref{MetricAnsatz}--\eqref{PhiAnsatz} using the recipe of \cite{Son:2002sd} and subsequently justified in \cite{Herzog:2002pc} starting from the more fundamental prescription of \cite{Gubser:1998bc,Witten:1998qj}.  In our case, the relevant metric perturbations exhibit rotational symmetry and thus only consist of $\delta g_{00}$, $\delta g_{ii}$, $\delta g_{55}$, $\delta g_{05}$, and $\delta \phi$.  Without loss of generality, we assume $\delta g_{11} = \delta g_{22} = \delta g_{33}$.  Henceforth we work in the gauge $r=\phi$, which is convenient because the equation for $\delta g_{11}$ decouples from the other perturbations.  Setting $h_{11} = e^{-2 A} \delta g_{11}$, we find that the equation for $\delta g_{11}$ reduces to
 \begin{align}\label{ViscosityEquation}
   h_{11}'' &= \left( -{1 \over 3A'} - 4A' + 3B' -
     {h' \over h} \right) h_{11}'  \nonumber \\
    &\qquad{} +
    \left( -{e^{-2A+2B} \over h^2} \omega^2 +
     {h' \over 6h A'} - {h' B' \over h} \right) h_{11} \,,
 \end{align}
where primes denote derivatives with respect to $\phi$.  We impose the normalization condition $h_{11} \approx 1$ at the boundary of AdS, as well as infalling boundary conditions at the black hole horizon:
 \begin{equation}
   h_{11} \approx c_{11}^{-} e^{i \omega t} \left|\phi-\phi_H \right|^{-i \omega/4 \pi T} \,.\label{HorizonAsymptotics}
 \end{equation}

At any value of $\phi$ between the horizon at $\phi = \phi_H$ and the conformal boundary at $\phi = 0$, the number flux of $h_{11}$ quanta with frequency $\omega$ falling into the black hole is given by the conserved quantity
 \begin{equation}
    {\cal F}(\omega) =  {e^{4 A - B} h \over 4 A'^2} \left| \Im h_{11}^* h_{11}' \right| \,. \label{Flux}
 \end{equation}
Given the number flux ${\cal F}(\omega)$, one can compute the imaginary part of the retarded two-point function of $T_{ii}$ from
 \begin{equation}
   \Im G_R(\omega) = -{2 {\cal F}(\omega) \over \kappa_5^2}\,.  \label{GRFromFlux}
 \end{equation}
Heuristically, \eqref{GRFromFlux} is the statement that dissipation in the boundary theory is related to the probability for particles from the conformal boundary of AdS to be absorbed into the horizon.

From \eqref{HorizonAsymptotics} and \eqref{Flux}, one straightforwardly finds
 \begin{equation}
  {\cal F} (\omega) = \omega e^{3 A(\phi_H)} {\left| c_{11}^{-} \right|^2 \over 4 A'(\phi_H)^2} \,,
 \end{equation}
which gives
 \begin{equation}
    {\zeta \over s} = {1\over 4 \pi} \left| c_{11}^{-} \right|^2 {V'(\phi_H)^2 \over V(\phi_H)^2}\,. \label{ZetaOverS}
 \end{equation}
In deriving \eqref{ZetaOverS} we have used the relation $A'(\phi_H) = - V(\phi_H)/3 V'(\phi_H)$, which follows from the equations of motion for the background \eqref{MetricAnsatz}--\eqref{PhiAnsatz}.  Bulk viscosity measures the hysteresis in nearly adiabatic $SO(3)$-invariant perturbations of the thermal medium.  Therefore, to extract $\zeta$, we can compute the quantity $c_{11}^-$ appearing in \eqref{ZetaOverS} in the $\omega \to 0$ limit.

A more detailed explanation of the derivation of \eqref{ZetaOverS}, as well as a description of the numerical computation of $c_{11}^-$, is given in \cite{GPR}. Figure~\ref{EOSandZetaComparison} includes a plot of $\zeta/s$ for the potentials \eqref{FirstPotential} and \eqref{TunedV}, as well as for pure glue as obtained in \cite{Meyer:2007dy} from lattice simulations, and for QCD as obtained from the sum rule approach of \cite{Kharzeev:2007wb,Karsch:2007jc}.

An alternative approach to constructing black holes that describe the thermal phase of non-conformal gauge theories has recently been suggested in \cite{Gursoy:2008bu}, following earlier work \cite{Gursoy:2007cb,Gursoy:2007er}.  The setup is similar to \eqref{Action}, except that $V(\phi) \sim -\phi^{1/2} e^{\sqrt{2 \over 3} \phi}$ at large $\phi$.  It is argued that fluctuations around the zero-temperature background (which is singular) yield a glueball spectrum with $m^2 \sim n$ for large excitation numbers $n$, suggestive of linear confinement.  The black hole solutions of \cite{Gursoy:2008bu} have a minimum temperature $T_{\rm min}$, where the specific heat diverges and $c_s^2$ vanishes.  The equation of state resembles that of pure Yang-Mills theory, with a sharp transition at some temperature $T_c$.  We will assume that $T_c = T_{\rm min}$, although it is more generic to have a first order transition at some $T_c > T_{\rm min}$.  The potentials employed in \cite{Gursoy:2007cb,Gursoy:2007er,Gursoy:2008bu} do not always have maxima, but the behavior of black holes close to $T_c$ is similar to what one finds with
 \begin{equation}\label{ConfinementPotential}
   V(\phi) = {-12 (1 + a \phi^2)^{1/4} \cosh \sqrt{2 \over 3} \phi +
     b \phi^2 \over L^2} \,.
 \end{equation}
Potentials of this form also exhibit a glueball spectrum with $m^2 \sim n$.

Because the phase transition is sharp for potentials of the form \eqref{ConfinementPotential}, one might expect the behavior of $\zeta/s$ near $T_c$ also to be sharp.  And so it proves, as shown in figure~\ref{EOSandZetaComparison}.  But $\zeta/s$ remains finite at $T_c$, even as the specific heat diverges, similar to behavior reported in \cite{Buchel:2007mf}.  Exploration of potentials similar to \eqref{ConfinementPotential} reveals the following related behaviors:
 \begin{itemize}
  \item When the potential \eqref{ConfinementPotential} is modified to match the equation of state of pure glue more closely, the peak in $\zeta/s$ becomes broader and lower, and $\zeta/s$ becomes bigger well away from $T_c$.
  \item It appears that $\zeta/s$ never diverges as long as $c_s^2\geq 0$, which is equivalent to positive specific heat in the absence of chemical potentials.  Potentials of the form \eqref{ConfinementPotential} exhibit solutions with $c_s^2<0$.  In some cases, $c_s^2 \to -\infty$, corresponding to a minimum of the entropy density, and then $\zeta/s$ does diverge.  Similar behavior arises when $V(\phi)$ has a narrow region of sharp decrease.  Of course, when $c_s^2<0$, the significance of $\zeta$ is more formal, because one is perturbing around an unstable background.
  \item When $V'(\phi)/V(\phi)$ is slowly varying, $\zeta/\eta \approx 2 ({1 \over 3} - c_s^2)$.  This adiabatic approximation can be derived by dimensional compactification of a conformal field theory, and it is generally a good indicator of the order of magnitude of $\zeta$, even when $c_s^2<0$.  See \cite{Benincasa:2005iv,Benincasa:2006ei} for related observations, and in particular \cite{Buchel:2007mf} for the conjecture that $\zeta/\eta \geq 2 ({1 \over 3} - c_s^2)$ for all black hole solutions.
 \end{itemize}

In conclusion: five-dimensional gravity coupled to a single scalar contains the minimum amount of freedom needed to match an equation of state to a family of black holes.  Constructions similar to those of \cite{Gursoy:2008bu} show even sharper behavior in the equation of state near $T_c$ than pure glue does.  These constructions lead to a sharp rise in $\zeta/s$ near $T_c$, though $\zeta/s$ remains finite at $T_c$.  The steepness of this rise is associated with the proximity of a minimum of $s$ on a branch of thermodynamically unstable solutions.  More generally, we conjecture that $\zeta/s$ diverges precisely when and where $s$ has an extremum \footnote{We intend this conjecture to apply even when $s$ is not differentiable as a function of $T$\@.  The closest one can get to having a divergent $\zeta/s$ on a thermodynamically stable branch of solutions is for such a branch to meet at an angle with a metastable branch in the $s$ versus $T$ plane.  Then a divergence in $\zeta/s$ will appear right at the point where the two branches meet.}.  Constructions of \cite{Gubser:2008ny}, in which it is assumed that the smooth but rapid cross-over of QCD is described in terms of smooth but rapid cross-over behavior of dual black hole solutions, lead to a more modest rise in $\zeta/s$ near $T_c$.

The steep rise of $\zeta/s$ observed for the potential \eqref{ConfinementPotential} is at least broadly consistent with the results of \cite{Kharzeev:2007wb,Meyer:2007dy}, although it is perhaps troubling that blunter behavior arises for potentials like \eqref{TunedV} that more closely match the pure glue equation of state.

The more modest rise of $\zeta/s$ observed for the potential \eqref{FirstPotential} and its variant \eqref{TunedV} indicates some tension with the results of \cite{Karsch:2007jc}.  Our results suggest that $\zeta/s \lesssim 0.1$ at $T_c$ for QCD\@.  If $\zeta/s$ is significantly bigger there, it either means that there is a subtlety in the matter lagrangian that we have not understood, or that Einstein gravity does not adequately describe the approach to confinement from above.  If, instead, our results are closer to the true behavior of real-world QCD, it suggests that the parametrization of the spectral function in \cite{Kharzeev:2007wb,Karsch:2007jc} is somehow misleading, and that the lattice study \cite{Meyer:2007dy} needs to be extended to include fermions before being directly compared with QCD close to $T_c$.

\begin{acknowledgments}
We thank D.~Kharzeev, H.~Meyer, and A.~Polyakov for useful discussions.  This work was supported in part by the Department of Energy under Grant No.\ DE-FG02-91ER40671 and by the NSF under award number PHY-0652782.  The work of A.N.~was also supported by the Office of Naval Research via an NDSEG Fellowship.  F.D.R.~was also supported in part by the FCT grant SFRH/BD/30374/2006.
\end{acknowledgments}

\bibliographystyle{apsrev}
\bibliography{sound}

\end{document}